\documentclass{jfm}
\usepackage{newtxtext}
\usepackage{newtxmath}
\usepackage{amsfonts} % collection of miscellaneous TeX fonts
\usepackage{amsmath} 
\usepackage{amssymb}
\usepackage{bm}% bold math
\usepackage{natbib}
\usepackage[section]{placeins}
\usepackage{graphicx}
\usepackage{hyperref}
\usepackage{booktabs}
\usepackage{subcaption}
\usepackage{siunitx}
\usepackage{float}
\usepackage{ulem}
\hypersetup{
	colorlinks = true,
	urlcolor   = blue,
	citecolor  = black,
}

\newcommand{\RomanNumeralCaps}[1]
\linenumbers
\title{Pair statistics of oblate spheroids settling in a turbulent flow}

\author{Prateek Anand\corresp{\email{prateekanand85@gmail.com}} and Samriddhi Sankar Ray}
\affiliation{International Centre for  Theoretical Sciences, Tata Institute of Fundamental Research,  Bangalore 560089, India}

\begin{document}

\maketitle

\begin{abstract}

	%The turbulent transport of anisotropic particles is more complex than
	%the motion of spherical, inertial particles because of the additional
	%degrees of freedom. Even for the simplest case of spheroids with
	%minimal anisotropy, the emergence of a gravity-induced torque leads to
	%a significant change in the way such particles orient as they settle
	%under gravity.  Such effects become more complex for an underlying
	%turbulent flow as is the typical setting for ice crystals in a cloud.
	%However, a detailed understanding of the effect of all these competing
	%interactions on small-scale, pair-statistics still remains an open
	%question. Thus, 

	We perform direct numerical simulations of sub-Kolmogorov, inertial
	spheroids settling under gravity in homogeneous, isotropic turbulence
	and find that small-scale clustering, measured via the correlation
	dimension, depends sensitively on the spheroid aspect ratio. In particular,
	such spheroids are shown to cluster more as their anisotropy increases.
	Further, the approach rate for pairs of spheroids are calculated and
	found to deviate significantly from the spherical-particle limit. Our
	study, spanning a range of Stokes numbers and aspect ratios, provides
	critical inputs for developing collision models to understand the
	dynamics of sedimenting, anisotropic particles in general and ice
	crystals in clouds in particular.

\end{abstract}

\section{Introduction}
%Recent studies have underlined the importance of the precise dynamics of ice crystals in cold clouds on our atmospheric radiation budget~\citep{liou1986}.
 
Cirrus clouds cover roughly $20\%$ of the earth's atmosphere and are a major component in weather and climate research since they affect the radiation field of the earth-atmosphere system~\citep{liou1986}. At a microphysical level, these clouds are composed of anisotropic ice crystals that scatter sunlight~\citep{baran2004}. Such clouds are necessarily turbulent~\citep{smith1996} and hence the basic mechanism --- that of a turbulent suspension of anisotropic particles --- at the heart of this problem is most usefully studied with tools of turbulent transport and statistical physics.  Of course, such suspensions are not unique to cold
clouds but seen in a wide variety of phenomena~\citep{voth}. These range from paper making
process involving cellulose fibres~\citep{lundell2011}, the use of biopolymers
in pharmaceuticals and cosmetics~\citep{erni2009}, soot particles in combustion
affecting the radiative forcing~\citep{moffet2009} to pollen
dispersion~\citep{sabban2011} and the dynamics of marine
microorganisms~\citep{pedley1992,SSRay2020-Flocking}. 

However despite its ubiquity, studies of turbulent suspensions of anisotropic
particles, such as spheroids, are fairly recent. This is especially true when
compared to the enormous progress made in our understanding of suspensions of
small, spherical particles~\citep{brandt2022,becreview}. Remarkably, the
understanding of such simpler suspensions have been successfully exploited to
answer specific and important questions related to warm clouds such as
collisions~\citep{mei1999,ayala2008effects,saw2014,ireland2016,PicardoCollisions},
coalescences~\citep{grabowski2013,bec2016,James17}, and settling of water
droplets~\citep{ssray2014,naso2018} --- all of which relate to the problem of rain
initiation~\citep{falkovich2002,shaw2003}.

Answering similar questions for cold clouds is harder. This is because the anisotropic particles have not only additional degrees of freedom but also
forces and torques which depend on their orientations. These, as we shall see
below, lead to a much more complicated set of equations of motion which need
more care in its numerical implementation and interpretation. Indeed, while
important results have emerged on the effects of anisotropy in turbulent flows,
these have largely been in the context where either the gravitational effects are neglected~\citep{Roy2018,SSRay2020-Flocking,Allende_Bec_2023} or the effects of gravity are not included consistently across all the aspects of particle dynamics~\citep{Pumir_2011,Gupta_2014,siewert2014,PumirPRL,JuchaPRF}.
%inertial~\citep{Pumir_2011,Gupta_2014,siewert2014,PumirPRL} or gravitational effects~\citep{Roy2018,SSRay2020-Flocking,Allende_Bec_2023} on the translational or rotational dynamics~\citep{PumirPRL,JuchaPRF} have been ignored. 
Matters are further complicated by observations of a wide variety of shapes and sizes that these crystals come in~\citep{klett1998}.

To correctly account for the motion of anisotropic particles settling in turbulence, one needs to first study the motion of the aforesaid particles settling in a quiescent ambient in the presence of small but finite fluid inertia. The earliest of these studies was done by~\cite{cox1965}, who calculated the gravity-induced torque to $O(Re_s)$ on a spheroid of small eccentricity, and later, \cite{khayatcox1989} obtained this torque for a slender body; here $Re_s$ is the Reynolds number based on the settling speed of the particle. The analogous calculation for a spheroid of arbitrary aspect ratio ---  a compelling model for ice crystals --- was performed only recently by~\cite{Dabade2015}. In all of the aforementioned scenarios, the gravity-induced torque acts to orient both prolate and oblate spheroids in a broadside-on orientation.
%Only recently, the effect of gravity on the orientation dynamics of spheroids settling in a quiescent ambient have been studied by~\cite{Dabade2015}, who found that small but finite fluid inertia on particle scales exerts a gravity-induced torque on the spheroids and orients them in the broadside-on orientation. 
When settling in a \textit{turbulent} flow, besides the gravity-induced torque, such spheroids are acted upon by a turbulent shear-induced torque. It is the competition between these two torques that gives rise to interesting orientation dynamics for settling spheroids as reported recently in ~\cite{sheikh2020,anand2020,gustavsson2021,sheikh2022}.
%Therefore, not surprisingly, it is only recently that a complete set of self-consistent equations for spheroids falling under gravity in --- a compelling model for ice crystals --- have been written down~\citep{Dabade2015} and numerically explored~\citep{anand2020} in the turbulent regime. 
In particular, such modelling has lead to an understanding of the origins of a broadside-on orientation~\citep{Gustavsson_2019} of ice crystals in turbulent Cirrus clouds. However, this is just one of the important characterisations for anisotropic particles, while fundamental questions of pair-statistics, involving collision and coalescence rates, remain largely unexplored.

In this paper we answer these questions and make a careful study of the effects
of particle shape and size in determining the nature of small-scale clustering
and the effective approach rates of colliding particles in a sedimenting,
turbulent suspension of spheroids. While not actually resolving collisions in
our system, akin to the \textit{ghost collision approach}~\citep{wang1998,ayala2008effects} used extensively for
spherical particles, the measurement of such approach rates form the building
blocks for modelling the collision kernels in ice-laden clouds~\citep{sheikh2022}. 

\section{Formulation}\label{sec:form}
We perform direct numerical simulations (DNSs) of a dilute
suspension of non-interacting spheroids sedimenting in a statistically
homogeneous, isotropic turbulent flow.  The fluid velocity field ${\bf u}$ satisfies the
incompressible Navier-Stokes equations and is driven to a statistically steady
turbulent state via an injection of constant energy at the lowest wave numbers.

For a given inertial spheroid, acted upon by gravitational acceleration  $g \hat{\bm{g}}$, the equations governing its translational 
$\bm{U}_p$ and angular $\bm{\omega}_p$ dynamics are given by
\begin{align}
	&\frac{d{\boldsymbol U}_p}{dt} =  g{\hat{\bm{g}}}+\frac{1}{\tau_p X_A}\bm{M}_t^{-1}\bm{\cdot}(\bm{v}-\bm{U}_p) \label{eq:translation}, \\
	&\frac{d\bm{\omega}_p}{dt}+\bm{I}_p^{-1}\bm{\cdot}[\bm{\omega}_p \bm{\wedge} (\bm{I}_p\bm{\cdot\omega}_p)] = 
	K_{sed} \bm{I}_p^{-1}\bm{\cdot}\left[(\hat{\bm{g}}\bm{\cdot p})\,(\hat{\bm{g}}\bm{\wedge p})\right] \nonumber \\
	&+8\pi\mu L^3 \bm{I}_p^{-1}\bm{\cdot}\left[{\bm M}_r^{-1}\bm{\cdot}\left(\frac{1}{2}\bm{\Omega}-\bm{\omega}_p\right)
	-Y_H (\bm{E\cdot p})\bm{\wedge p}\right],  \label{eq:rotation}
\end{align}
where $\bm{v}$ is the fluid velocity at particle position. The large particle-to-fluid density ratio ($\rho_p/\rho_f$), relevant to the atmospheric scenario, allows the neglect of Basset and added mass forces in eq.~\eqref{eq:translation}. Particle inertia is characterized by a Stokes number defined as $St = \tau_p/\tau_\eta$, the ratio of the particle relaxation time $\tau_p$ (defined below) and the characteristic turbulent (Kolmogorov) time scale $\tau_\eta$. The particle response time $\tau_p$
(and hence the Stokes numbers) also depend sensitively on the spheroid aspect ratio
$\kappa$~\citep{anand2020}; thus different values of $\kappa$ lead to different
ranges of the Stokes numbers for our particles. Here, $\kappa=b/L<1$ for oblate spheroids, where $b$ and $L$ are the semi-axis lengths along and orthogonal to the symmetry axis $\bm{p}$, respectively. The angular dynamics of the settling spheroids also depend sensitively on
the moment of inertia tensor $\bm{I}_p$ and the Stokesian translational
($\bm{M}^t$) and rotational ($\bm{M}^r$) mobility tensors, with the latter two ensuring an effective instantaneous drag and torque on the spheroids~\citep{kim2013}. The mobility tensors are defined as: $\bm{M}_{t(r)}=X_{A(C)}^{-1}(\kappa)\bm{pp}+Y_{A(C)}^{-1}(\kappa)(\bm{I-pp})$~\citep{kim2013}, where the principal resistance coefficients ($X_A-Y_C$) are functions of only the spheroid aspect ratio $\kappa$ and $\tau_p=2\rho_p L^2 \kappa$/(9$\rho_f \nu X_A$), with $\nu$ being the kinematic viscosity.

For an advecting turbulent flow with kinematic viscosity $\nu$ and mean dissipation rate
$\epsilon$, the Kolmogorov shear rate $\dot{\gamma}_\eta=\tau_\eta^{-1}=(\epsilon/\nu)^{1/2}$
allows us to define a particle Reynolds number
$Re_{\dot{\gamma}_\eta}=\dot{\gamma}_\eta L^2/\nu < 1$ (by definition for $L\ll\eta$); likewise another particle Reynolds number $Re_s=U_s L/\nu$ can be defined based on the slip velocity $U_s=\tau_p g$. We limit ourselves to small but finite values of $Re_s$, which ensures that the dominant forces acting on our
spheroids are the gravitational and the quasi-steady Stokesian drag
forces~\citep{Jeffery,kim2013}. Further, considering small (sub-Kolmogorov)
particles allows us to define the turbulent torque via the Bretherton's constant
$Y_H/Y_C=(\kappa^2-1)/(\kappa^2+1)$~\citep{Bretherton_1962}. This turbulent
torque competes with the gravity-induced torque~\citep{Dabade2015} and
the relative importance of one to the other is quantified via the ratio
$\mathcal{T} = K_\text{sed}/\mu L^3 \dot{\gamma}_\eta\sim U^2 f_I(\kappa)$~\citep{sheikh2020,anand2020,gustavsson2021},
where $U=U_s/u_\eta$ ($u_\eta$ is the
characteristic, small-scale Kolmogorov velocity field) and 
$f_I(\kappa)$, which characterizes the gravity-induced torque by setting $K_{sed}=Re_s\mu U_s
L^2f_I(\kappa)X_A/Y_A$ in \eqref{eq:rotation}, was obtained in closed-form by \cite{Dabade2015} for a spheroid of arbitrary aspect ratio $\kappa$. Finally, the effects of the advecting
turbulent flow enter the rotational dynamics of the spheroids via the vorticity
$\bm{\Omega}$ and rate-of-strain $\bm{E}$ fields evaluated simultaneously from
the incompressible, Navier-Stokes equation. The fluid characteristic
small-scale acceleration $a_\eta=u_\eta/\tau_\eta$ also allows us to define a Froude number $Fr
= a_\eta/g$ to measure the effects of the turbulent acceleration with respect
to that set by gravity; in what follows we simulate for $Fr = 0.05$, comparable
to typical settings in clouds~\citep{shaw2003,ssray2014}, and as a contrast also for the zero gravity case
which leads, by definition, to $Fr = \infty$.

Our DNSs involve simultaneous simulations of the turbulent fluid as well as the
time integration of spheroid trajectories defined by
Eqs.~\eqref{eq:translation} and \eqref{eq:rotation}. The turbulent flow is
simulated by using a standard pseudospectral method with a second-order
Adams-Bashforth scheme for time-marching. A spatial resolution of $512^3$
collocation points ensures a Taylor-scale based Reynolds number
$Re_\lambda=150$. A $2/3^\text{rd}$ - dealiasing rule is used in the pseudo-spectral method, resulting in $k_\text{max}\eta\approx 1.5$. A trilinear interpolation scheme is used to obtain the fluid
velocity field at the particle positions for solving
Eq.~\eqref{eq:translation} (see Ref. \cite{Ray2018} for details). The set of parameters used in our simulations 
are given in Table~\ref{tab:parameter_list}. We use
different values of aspect ratios $0.01 \leq \kappa \leq 0.95$ (thus
restricting ourselves to oblates) and Stokes numbers $0 \leq St \leq 1.1$,
where $St = 0$ is the tracer limit. To obtain reliable statistics, we use $10^6$ oblates for every combination of $\kappa$ and $St$. Unlike many earlier studies, we have incorporated a gravity-induced torque (as explained earlier) in our model for 
anisotropic particles. This necessarily constraints our model to be accurate for particle Reynolds numbers $Re_s$ less than unity; 
in our simulations we ensure $Re_s=0.5$. It is instructive to rewrite the particle Stokes number in terms of the particle Reynolds 
number $Re_s$, the density ratio  $\rho_p/\rho_f$ and the aspect ratio $\kappa$ as 
$St=\tau_\eta^{-1}(2\rho_p\kappa\nu Re_s^2/(9\rho_f X_A g^2))^{1/3}$. This illustrates clearly that while, in principle, we 
could chose any value of $St$, for the model to be physically realisable, the range of Stokes numbers are restricted (see Table Table \ref{tab:parameter_list}). 
Nevertheless, we do consider a wide range of parameters such as the torque ratio $\mathcal{T}$ and the density ratio $\rho_p/\rho_f$. 
We also alert the reader that as $\rho_p/\rho_f\sim O(1)$, some of the simplifications (such as the neglect of the Basset history and the added mass forces) in 
\eqref{eq:translation} is questionable. Nevertheless for completeness, we do simulate (for every aspect ratio) a set of particles 
with $\rho_p/\rho_f=1$ (see Table \ref{tab:parameter_list}); but the remaining bulk of our results pertain to cases where the density ratios are comfortably larger than unity. %That being said, the Stokes numbers do vary a little from one aspect ratio to the other.

\begin{table}
	\begin{center}
		\def~{\hphantom{0}}
		\begin{tabular}{lcccc}
			$\kappa$&$\rho_p/\rho_f$  & $St$   &   $U$ & $\mathcal{T}$ \\[3pt]
			&10000& 0.254 & ~~5.08~ & 61\\
			&~5000& 0.201 & ~4.03 & 38\\
			&~1000& 0.118 & ~2.36 & 13\\
			&~~500& 0.094 & ~1.87 & 8\\
			0.01~ &~~100& 0.055 & ~1.09 & 3\\
			&~~~50& 0.043 & ~0.87 & 2\\
			&~~~10& 0.025 & ~0.51 & 0.6\\
			&~~~~5& 0.02 & ~0.4 & 0.4\\
			&~~~~1& 0.012 & ~0.24 & 0.1\\
			\hline
			&10000& 0.546 & ~~10.92~ & 270\\
			&~5000& 0.433 & ~8.67 & 170\\
			&~1000& 0.253 & ~5.07 & 58\\
			&~~500& 0.201 & ~4.02 & 36\\
			0.1~ &~~100& 2.35 & ~1.09 & 12\\
			&~~~50& 0.093 & ~1.87 & 8\\
			&~~~10& 0.054 & ~1.09 & 2.7\\
			&~~~~5& 0.043 & ~0.87 & 1.7\\
			&~~~~1& 0.025 & ~0.51 & 0.6\\
			\hline
			&10000& 0.915 & ~~18.31~ & 550\\
			&~5000& 0.726 & ~14.53 & 345\\
			&~1000& 0.425 & ~8.5 & 118\\
			&~~500& 0.337 & ~6.74 & 75\\
			0.5~ &~~100& 0.197 & ~3.94 & 25\\
			&~~~50& 0.156 & ~3.13 & 16\\
			&~~~10& 0.092 & ~1.83 & 5.5\\
			&~~~~5& 0.073 & ~1.45 & 3.5\\
			&~~~~1& 0.042 & ~0.85 & 1.2\\
			\hline
			&10000& 1.1 & ~~22.00~ & 107\\
			&~5000& 0.873 & ~17.47 & 67\\
			&~1000& 0.51 & ~10.22 & 23\\
			&~~500& 0.405 & ~8.11 & 14\\
			0.95~ &~~100& 0.237 & ~4.74 & 5\\
			&~~~50& 0.188 & ~3.76 & 3\\
			&~~~10& 0.11 & ~2.2 & 1.1\\
			&~~~~5& 0.087 & ~1.75 & 0.7\\
			&~~~~1& 0.051 & ~1.02 & 0.2\\
		\end{tabular}
		\caption{List of parameters used in the simulations}
		\label{tab:parameter_list}
	\end{center}
\end{table}

\section{Results}
\begin{figure}
	\includegraphics[width=1.0\columnwidth]{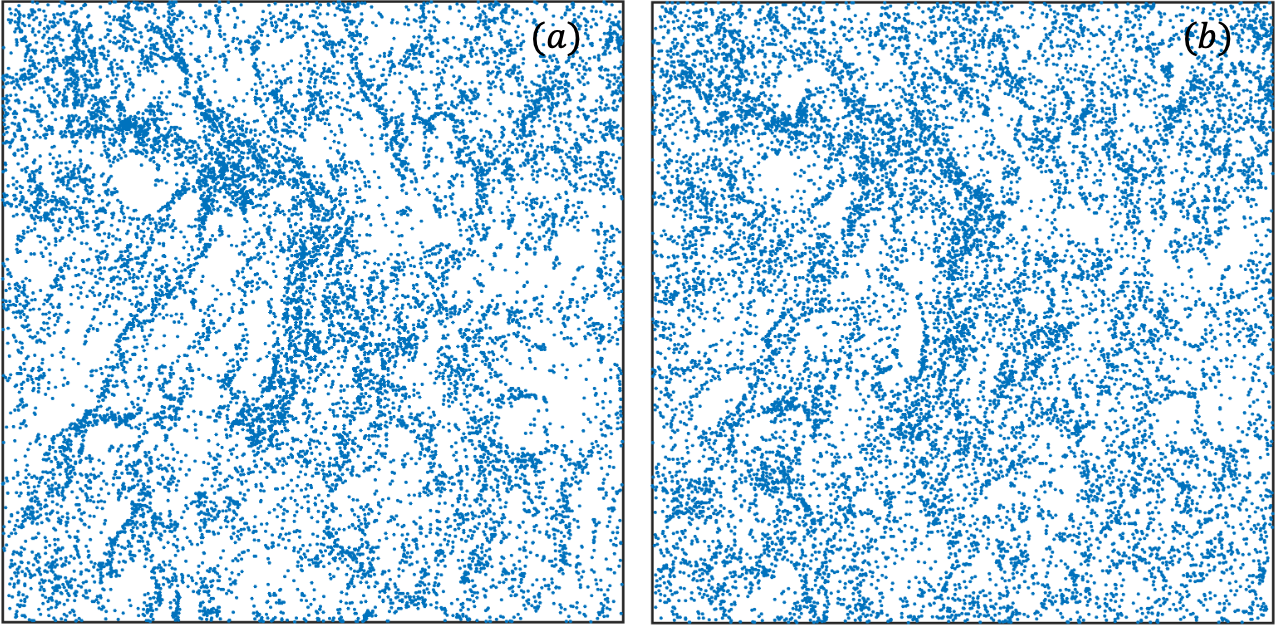}
	\caption{Representative snapshots, for $Fr = 0.05$, of particle position with comparable Stokes numbers ($St \sim 0.5$) but aspect ratios 
	(a) $\kappa = 0.1$ and (b) $\kappa = 0.95$; the (near) two-dimensional slice, on the $XZ$ plane, has dimensions $2\pi\times10\eta\times2\pi$. 
	A comparison of the two panels gives a visual cue that the disk-like ($\kappa = 0.1$) particles tend to cluster more strongly than the 
	near-sphere ($\kappa = 0.95$) ones.}
	\label{fig:Snapshots}
\end{figure}

\subsection{Small-scale clustering}
One of the striking effects in suspensions of inertial, spherical particles is
that of preferential concentration: particles tend to cluster, leading to a
non-homogeneous spatial distribution of the particles~\citep{maxey1993}, and the degree of clustering
depends on their Stokes number in a non-monotonous
way~\citep{Bec_frac_clustering,bec2007}. This degree of small-scale clustering is best
characterised by the correlation dimension $D_2$~\citep{bec2007} defined via the probability density
$p_2(r) \sim r^{D_2-1}$ of finding two particles at a distance $r \lesssim
\eta$, with $\eta$ being the characteristic small-scale (Kolmogorov scale) of
the flow. In the tracer $St \to 0$ and large $St \to \infty$ limits, the
correlation dimension $D_2 \to d$, where $d$ is the spatial dimension of the
problem; for finite Stokes numbers $D_2 < d$ with a minimum for $St \sim
O(1)$. The $D_2$ vs $St$ behaviour undergoes significant changes when
the gravitational accelerations become comparable~\citep{ssray2014} or larger
than the turbulent accelerations of the flow leading to important changes in
the collision rates for gravitationally settling, inertial spherical particles. 

The question of small-scale clustering for spheroids settling in a turbulent
flow is an open one. And yet, just like the case of spherical particles, such
clustering plays a key role in determining the eventual collision efficiency of,
for example, ice crystals in a cloud. It would be reasonable to expect that
the clustering of spheroids depends on their aspect ratio, and this expectation
is borne out in Figures~\ref{fig:Snapshots}(a) and (b), wherein particle
positions are plotted in a thin slice of the computational domain for a given
Stokes number. The extreme-shaped spheroids ($\kappa=0.1$) seem to cluster more
(notice the bigger voids) as compared to the nearly spherical particles
($\kappa=0.95$). The snapshots of particle clustering are suggestive at best and 
certainly not conclusive. Hence we estimate a precise measure of the degree 
of clustering in what follows.

\begin{figure}
	\includegraphics[width=1.0\linewidth]{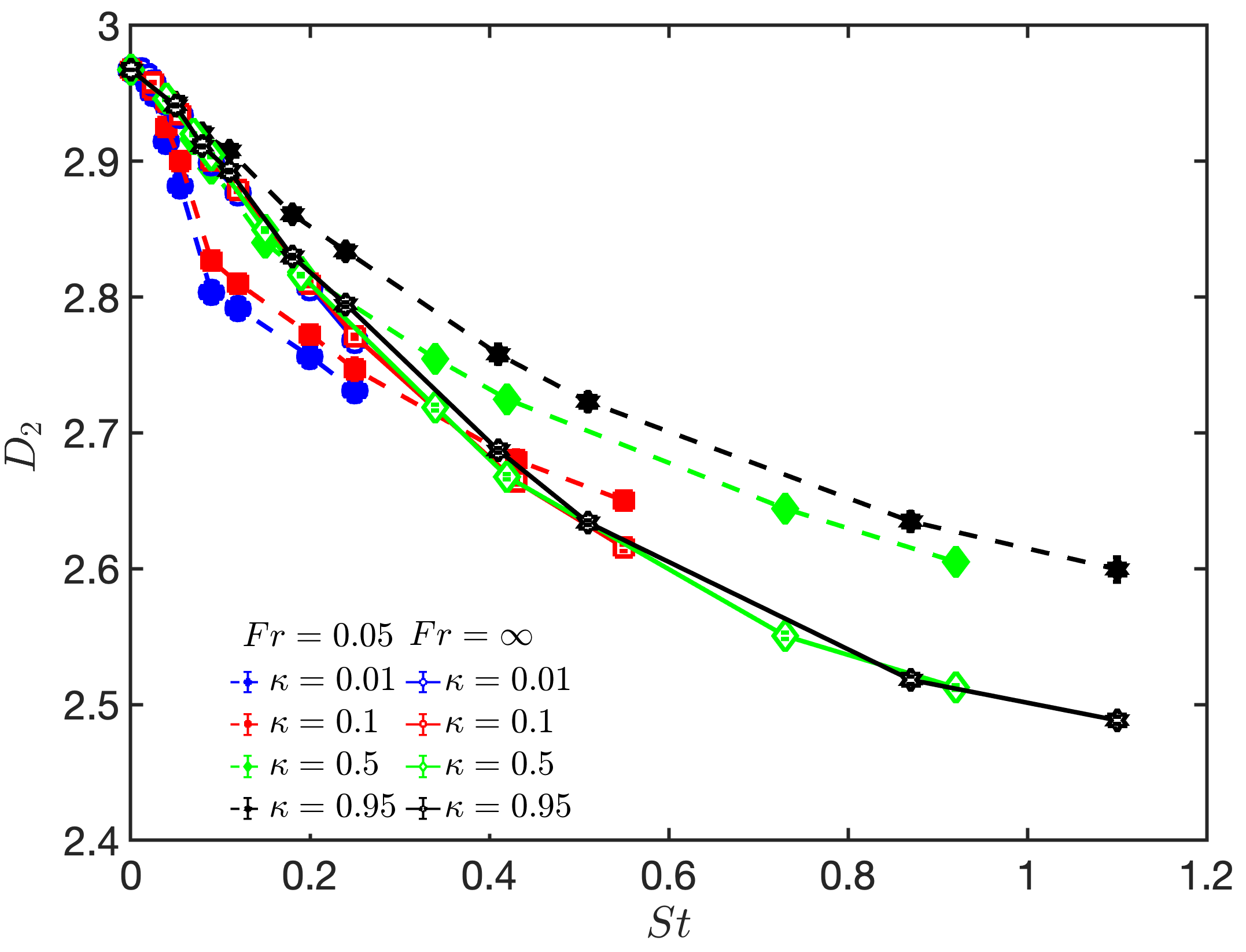}
	\caption{Correlation dimension $D_2$ for the spheroids as a function of
	their Stokes numbers, for different aspect ratios (see legend; filled
	symbols for $Fr=0.05$). Clearly, disk-like particles cluster more effectively, under
	gravity, than the near-sphere ones. The effect of gravitational
	settling on clustering is brought out by comparing the analogous plot
	for the zero gravity case (open symbols; $Fr = \infty$, see legend):
	anisotropy has no effect on clustering when gravity is absent.}
	\label{fig:D2vsStokes}
\end{figure}

To quantify our observations regarding particle clustering, we measure the
correlation dimension $D_2$ vs the Stokes number $St$ for different
\textit{kinds} of oblates as characterised by their aspect ratio $\kappa$. In
Figure~\ref{fig:D2vsStokes} we show a plot of the correlation dimension for
different oblates (dashed curves for $Fr=0.05$).  We observe that as spheroids become
disk-like $\kappa \to 0$, the degree of clustering becomes significantly
stronger compared to the spherical limit $\kappa \to 1$, which is consistent
with the observations of the particle concentration field in
Figure~\ref{fig:Snapshots}(a) and (b).

Note that the Stokes range for different values of $\kappa$ is different. This is because for
most systems of interest, the particle-to-fluid density ratio varies from
$1000$ (for instance, ice crystals in cold clouds) to $\sim10000$ for aerosols~\citep{bec2005}. Since the Stokes numbers are aspect ratio-dependent (see the expression in $\S$\ref{sec:form}),
$\rho_p/\rho_f=10000$ corresponds to different (maximum) Stokes numbers for each aspect ratio. 

\begin{figure}
	\includegraphics[width=1.0\columnwidth]{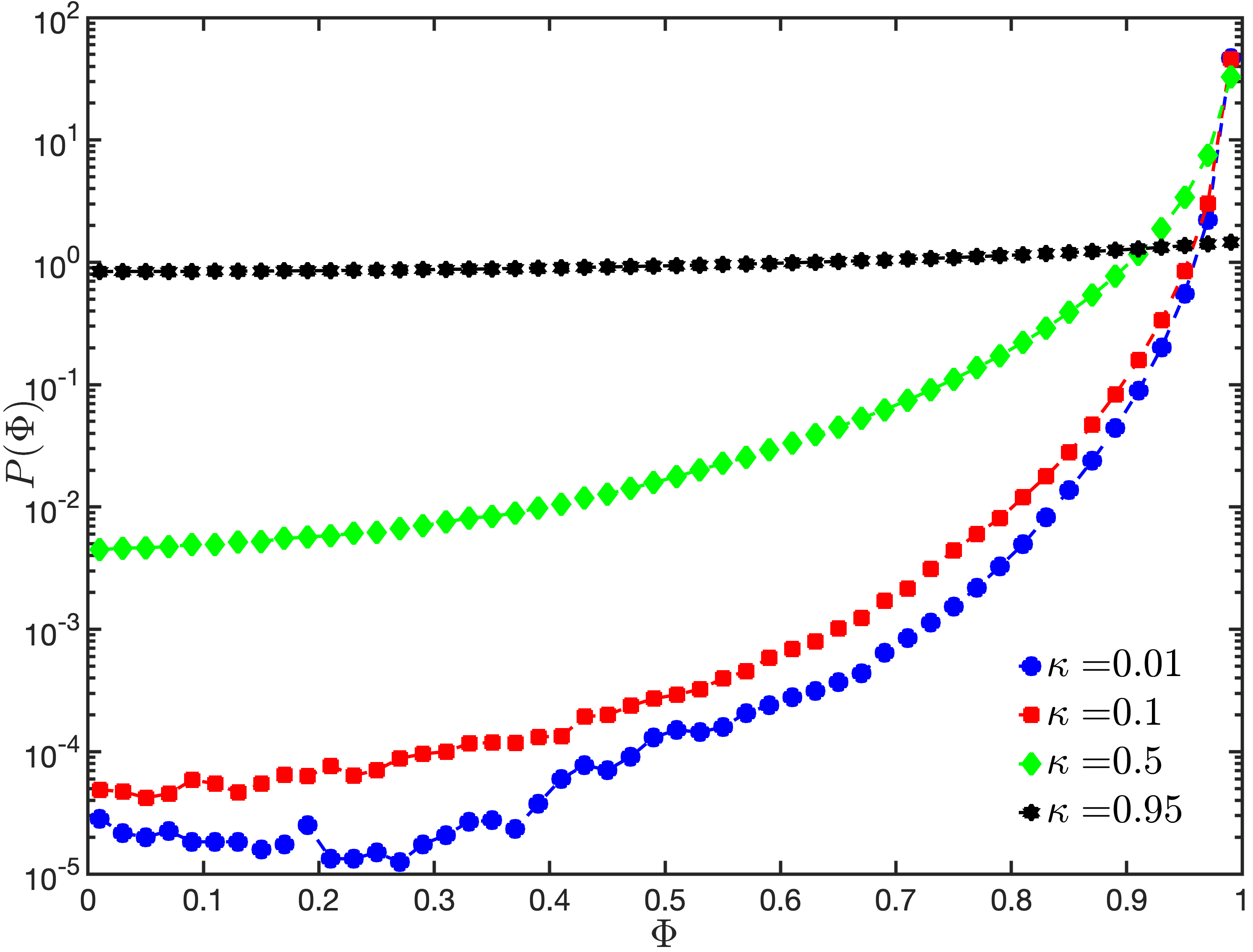}
	\caption{Distribution of the relative orientation $\Phi$ of the settling spheroids (with $St \sim 0.2$), 
	conditioned on their separation $r \leq 20\eta$, for different aspect ratios. %Clearly disk-like spheroids preferentially orient leading to a strong peak of the distribution as $\Phi \to 1$. 
	The ratio $\mathcal{T}\sim38,36,25$ and $3$ for $\kappa=0.01,0.1,0.5$ and $0.95$, respectively. Large values of $\mathcal{T}$ lead to a distribution that is sharply peaked at $\Phi\to1$ for disk-like spheroids. As particles become more isotropic, this bias weakens  eventually leading to a flat distribution in the near-sphere limit ($\kappa = 0.95$).}
	\label{fig:p1dotp2}
\end{figure}

Owing to the anisotropy of the spheroids, gravity plays an important role by
exerting the gravity-induced torque on the spheroidal particles. So does this
mean that in the absence of gravity, there is no aspect ratio dependence on the small-scale clustering? To
answer this question, we simulate the motion of the oblates for $Fr=\infty$ (no
gravity). In Figure~\ref{fig:D2vsStokes}, we show the plot of the correlation
dimension in the zero gravity case (solid curves). Clearly, and as anticipated
above, we find no dependence of $D_2$ on $\kappa$. In particular, we find the correlation dimension for the spheroids of $\kappa=0.95$ consistent with those obtained for settling spherical particles~\citep{ssray2014}.

The correlation dimension is a measure of the spatial organisation of the
particles and in Figure~\ref{fig:D2vsStokes} we have shown the importance of shape and gravity
in determining the eventual degree of clustering even if the Stokes numbers are
comparable.  However, the dynamics of the spheroids are not determined only by their
positional but also by their orientational information. To understand this, we
calculate the relative orientation (for the gravitationally settling spheroids)
$\Phi \equiv |\bm{p}_i\cdot\bm{p}_j|$ of pairs $i,j$ of spheroids conditioned on their separation $r$. 
Similar measurements have been obtained for suspensions of anisotropic particles, albeit in situations 
where the role of the gravity-induced torque has been neglected~\citep{siewert2014collision,JuchaPRF}. 
In our model of spheroids, we are able to investigate this more fundamentally as the orientation of individual spheroids is 
an outcome of the competition between torques 
resulting from gravity and turbulent shear, quantified through their ratio $\mathcal{T}$.
\begin{figure}
	\includegraphics[width=1.0\columnwidth]{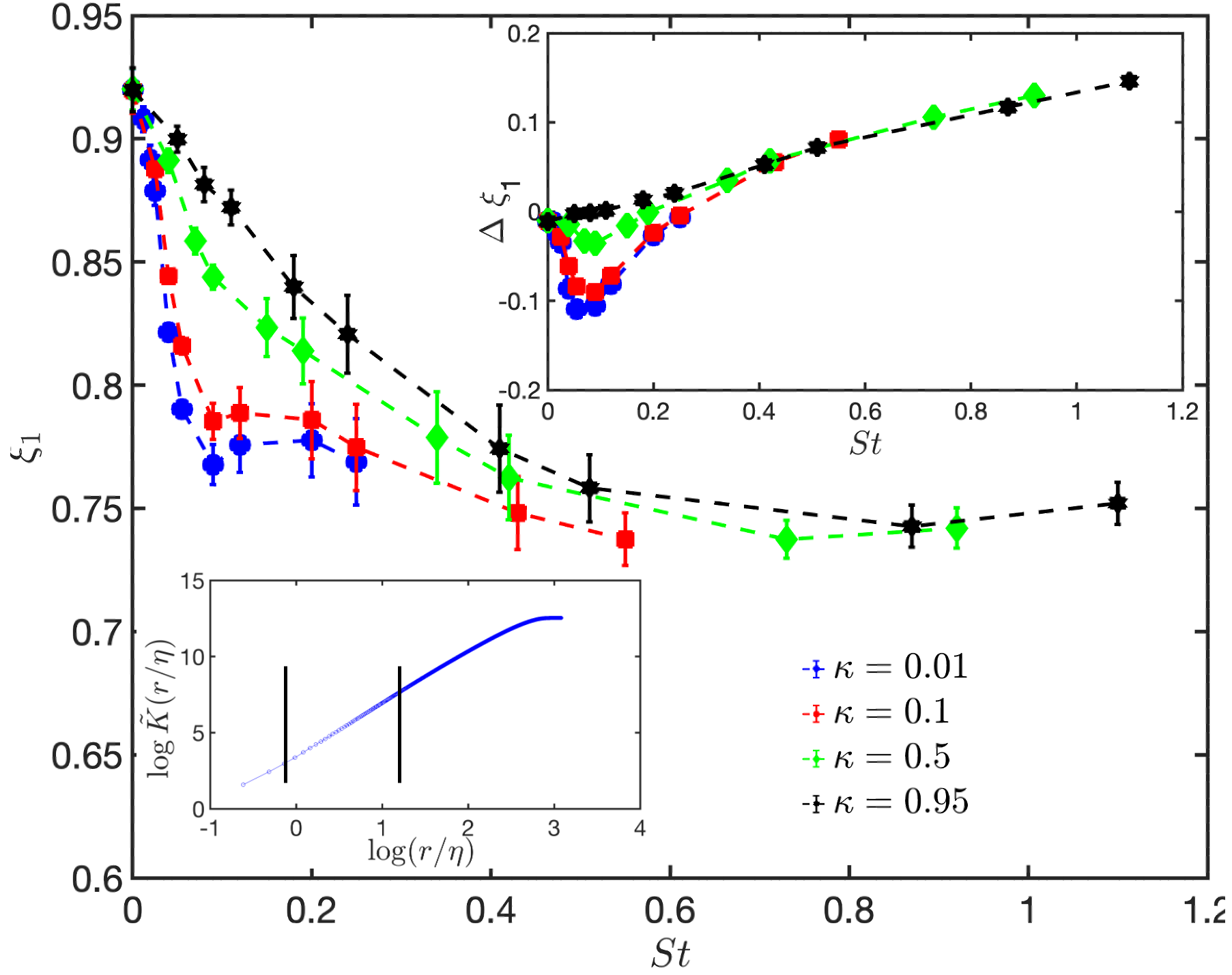}
	\caption{Variation of the exponent $\xi_1 = \gamma - D_2 + 1$, characterizing the relative longitudinal velocity for approaching, 
	sedimenting particles, as a function of the Stokes number and for different values of $\kappa$. In the upper inset, the difference $\Delta \xi_1 \equiv 
	\xi_1 (Fr = 0.05) - \xi_1 (Fr = \infty) $ between 
	this exponent for the gravity-driven settling and the zero-gravity case is shown. The lower inset shows a representative log-log plot of the cumulative sum $\tilde{K}(r)$ for $\kappa=0.5$ and $St=0.4$; the scaling 
	range, indicated by the pair of thick vertical lines, extend over a decade and yields $\gamma + 1$ and, thence, 
	the exponent $\xi_1$ (see text).}
	\label{fig:xiVsStokes}
\end{figure}
Na\"ively, one would expect for a given Stokes and large $\mathcal{T}$ %(whence $\kappa \to 0$), corresponding to disk-like spheroids, 
that the effects of gravity will dominate over
turbulent fluctuations leading to a distribution for $\Phi$ strongly peaked at
1, corresponding to a near perfect alignment, or orientational order, for
sedimenting oblates. %In the near-sphere limit of $\kappa \to 1$, 
On the other hand, when
$\mathcal{T} \sim O(1)$, %the lack of strong anisotropy 
the gravity-induced torque loses its dominance thereby ensuring that %nearly spherical 
the spheroids settle in arbitrary orientations leading to a more
even sampling of $\Phi$. In Figure~\ref{fig:p1dotp2}, we show measurements of $\Phi$ for $r \leq 20\eta$ and 
different values of $\kappa$; the Stokes number for all spheroids are close to 1. As our phenomenological 
picture suggests, the relative orientation is strongly peaked at $\Phi = 1$ for disk-like oblates $\kappa = 0.01$~($\mathcal{T}\gg1$) while becoming 
shallower as $\kappa$ increases and eventually becomes uniform for the near-spheres $\kappa = 0.95$~($\mathcal{T}\sim O(1)$). A similar behaviour is observed 
for other values of $St$ and $r$ (not shown). %The non-gravity case is compared, in the inset of Fig.~\ref{fig:p1dotp2}, to underline  once more the critical role of the gravity-induced torque on these particles.
\subsection{Relative longitudinal velocity}
Having investigated the orientational order and positional correlation for
settling spheroids, let us turn to the question of how such particles collide
with each other. This of course is the critical question in developing ideas of
collision kernels and estimating collision efficiencies in such systems and
particularly so in studies of ice crystals in cold clouds. There have been
several studies which have estimated the collision kernel and its dependence on
various parameters --- such as the aspect ratio of particles and the fluid
Reynolds numbers --- in earlier
studies~\citep{siewert2014collision,JuchaPRF,sheikh2022}. In this work, we take a slightly 
different point of view and address how violent the 
collisions can be through the approach rates~\citep{bec2005} of colliding
particles.

The approach rate can be
expressed as: $K(r)\equiv -\langle w\,\mathcal{H}(-w)|R=r\rangle p_2(r) \sim r^\gamma$
\citep{ssray2014}, where $w$ is the longitudinal velocity difference between
two particles and $\mathcal{H}$ is the Heaviside function. As is usual, instead of 
the exponent $\gamma$, we extract $\gamma + 1$ from 
loglog plots of the cumulative sum $\tilde{K}(r) \equiv \int_0^r K(r)$. Such cumulative sums 
allow a better scaling range and a representative plot of $\tilde{K}(r)$ is  
shown in the lower inset of Figure~\ref{fig:xiVsStokes} for $St = 0.4$ and $\kappa = 0.5$. 
Our scaling range, which is indicated by thick vertical lines, extend for over a decade and 
yield $\gamma + 1$ through a least-square fit on this range.
Following from
\cite{bec2005,bec2007,ssray2014}, when $r\ll\eta$, the approach velocity
$\langle w\,\mathcal{H}(-w)|R=r\rangle\sim r^{\xi_1}$, the approach rate
$K(r)\sim r^\gamma$ and $p_2\sim r^{D_2-1}$, which implies that
$\xi_1=\gamma-D_2+1$. Thus, our measured $\gamma + 1$ and $D_2$ allows an easy estimation 
of $\xi_1$.

The exponent $\xi_1$ characterizes the longitudinal
velocity differences between pairs of settling spheroids, and is plotted for
various Stokes numbers and aspect ratios in Figure~\ref{fig:xiVsStokes}. For 
nearly spherical particles, $\xi_1$ shows the same dependence with Stokes
number as reported for spheres in \cite{ssray2014}. As the aspect ratio
decreases, the constant-$\kappa$ curves start to get steeper. This implies that
for a given Stokes number, the velocity difference increases with
decreasing $\kappa$ or particles of smaller aspect ratio approach more
violently. The exponents reported in Figures \ref{fig:D2vsStokes} and \ref{fig:xiVsStokes} as well as the distribution of the relative orientations plotted in Figure \ref{fig:p1dotp2} are conditioned on $r\leq 20\eta$. Although not shown, we have confirmed that the statistics reported in this manuscript are consistent for smaller interparticle separations, such as, $r\leq5\eta$.
To underline the interplay between gravity and anisotropy, we plot
the difference between the $\xi_1$'s for $Fr=0.05$ and $Fr=\infty$ (zero
gravity) in the inset of Figure~\ref{fig:xiVsStokes} for various aspect ratios.
As evident from the inset, when $St\lesssim 0.05$, $\Delta\xi_1<0$ for
$\kappa=0.95$, implying that gravity strengthens the longitudinal velocity
difference of approaching spheroids. As $St$ increases beyond a critical value,
$\Delta\xi_1$ becomes positive which is indicative of gravity weakening the
velocity differences. This is consistent with the results reported in
\cite{ssray2014} for spherical particles. Interestingly, as the aspect ratio
decreases, $\Delta\xi_1$ starts behaving non-monotonically with $St$. It first
decreases and attains negative values before increasing and eventually
collapsing onto the near-spherical curve beyond a $\kappa$-dependent $St$. This
indicates that for $\kappa\leq0.5$, gravity first strengthens the velocity
difference with increasing $St$ up to $\sim0.08$. Thereafter the
gravity-induced approach velocity starts to reduce although $\Delta\xi_1$ still
remains negative; the minima of $\Delta\xi_1$ and therefore the gravity-induced
enhancement in velocity differences is sensitive to the aspect ratio of the
spheroid.  Beyond a $\kappa$-dependent Stokes number, $\Delta\xi_1$ attains
positive values thereby indicating that gravity now weakens the velocity
differences instead of enhancing it. Once all the curves coincide, the
difference of the longitudinal velocity behaves like those for spherical
particles. It is important to flag a technical point regarding the particle Reynolds number based on the Kolmogorov shear rate, $Re_{\dot{\gamma}_\eta}=\dot{\gamma}_\eta L^2/\nu$. Since the particles in our simulations are sub-Kolmogorov, $Re_{\dot{\gamma}_\eta}$ turns out to be quite small ($<0.1$) for most of the cases considered here. Therefore, the inertial correction to the turbulent shear-induced torque is sub-dominant in the present study for both $Fr=0.05$ and $\infty$. For the gravitational case, an order-of-magnitude comparison of the gravity-induced torque vis-a-vis the inertial correction to the turbulent shear-induced torque~$\sim U^2$, which is larger than unity for most of the cases in the present study~(Table~\ref{tab:parameter_list}).

\subsection{Alignment of orientation vectors with local vorticity and strainrate eigenframe}
\begin{figure} \centering \begin{subfigure}[b]{0.49\textwidth}
\includegraphics[width=\textwidth]{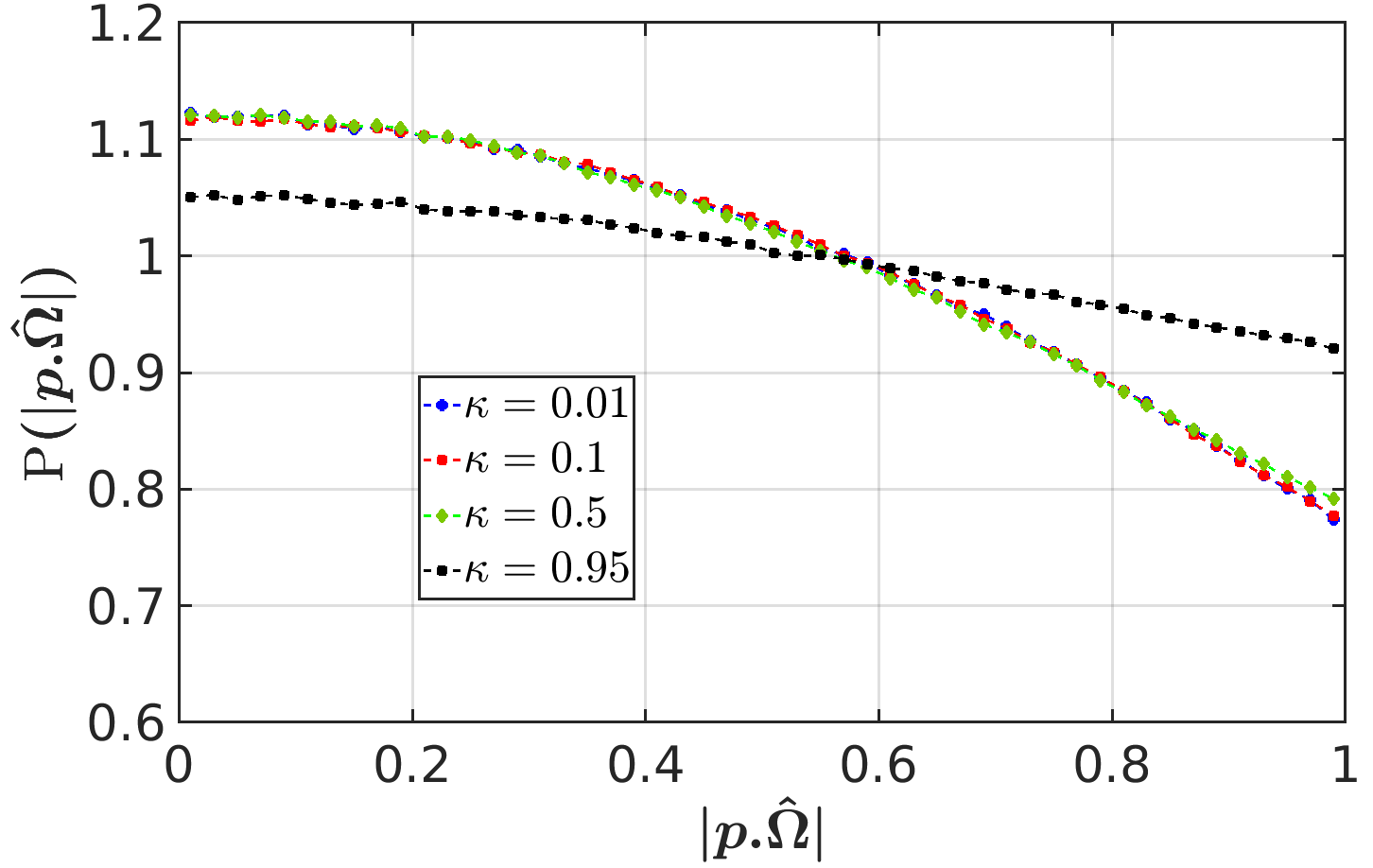} \caption{}
\end{subfigure} \hfill \begin{subfigure}[b]{0.48\textwidth}
	\includegraphics[width=\textwidth]{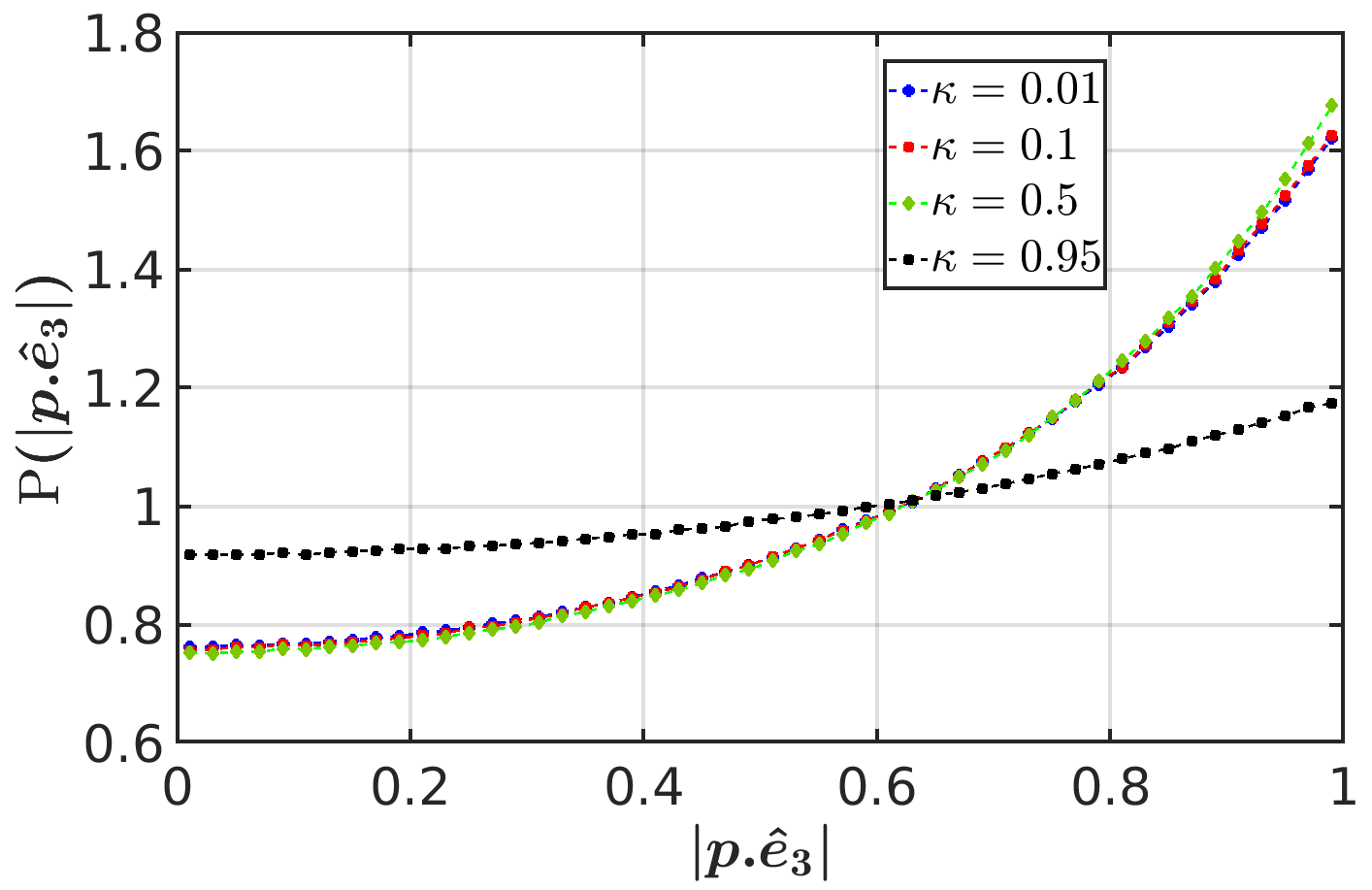} \caption{}
\end{subfigure} \caption{Probability distribution for a)
$|\bm{p\cdot}\hat{\bm{\Omega}}|$, where $\hat{\bm{\Omega}}$ is the local
vorticity direction, and b) $|\bm{p\cdot}\hat{\bm{e}}_3|$ where
$\hat{\bm{e}}_3$ is the direction of strongest contraction, obtained from DNS
for $St\approx0.2$ for various aspect ratios. The spheroid axis is weakly
perpendicular to the vorticity while strongly aligning with $\hat{\bm{e}}_3$.}
\label{fig:pdfpdot} \end{figure}

To augment our results on the pair statistics of oblate spheroids, we make a final calculation 
of how such particles align with the direction of 
the local vorticity $\hat{\bm{\Omega}}$ and the eigendirections of
the local rate of strain tensor. Figure \ref{fig:pdfpdot}a shows the normalized
probability distributions of the absolute value (due to the $\bm{p}\rightarrow
-\bm{p}$ symmetry) of the cosine of the angle between the spheroid axis
$\bm{p}$ and the local vorticity $\hat{\bm{\Omega}}$, for various aspect ratios
and $St \approx 0.2$. The distributions peak at
$|\bm{p\cdot}\hat{\bm{\Omega}}|=1$ for all $\kappa$ indicating that
vorticity lies along the largest plane of the oblate spheroids, leading to a tumbling motion for the oblate spheroids. Such a behaviour was already 
observed for tracer ($St=0$) spheroids
in~\cite{chevillard2013}. However, for our more complex anisotropic, inertial particles settling under gravity, 
the competing interactions of the additional torques (and forces) lead to a relatively weaker alignment than the tracer 
case studied by~\cite{chevillard2013}~(we have confirmed that for $St = 0$ we recover the results 
reported in~\cite{chevillard2013}.)

%\begin{figure} \centering
%\includegraphics[width=1.0\columnwidth]{pdoteomega_St_p2.png}
%\caption{Probability distribution of the absolute value of
%$\bm{p\cdot}\hat{\bm{\omega}}$, where $\hat{\bm{\omega}}$ is the local
%vorticity direction, obtained from DNS for $St\approx0.2$ for various aspect
%ratios. The spheroid axis is `weakly' perpendicular to the vorticity.}
%\label{fig:pdfpdotomega} \end{figure}

What about the preferential alignment with the local strainrate 
eigenframe $(\hat{\bm{e}}_1,\hat{\bm{e}}_2,\hat{\bm{e}}_3)$, where
$\hat{\bm{e}}_1$ is the direction of the strongest extension~(largest positive
eigenvalue), $\hat{\bm{e}}_3$ is the direction of strongest
contraction~(largest negative eigenvalue) and $\hat{\bm{e}}_2$ is the direction
corresponding to the intermediate eigenvalue, which can be either positive or
negative? The distributions for 
$|\bm{p\cdot}\hat{\bm{e}}_1|$ and $|\bm{p\cdot}\hat{\bm{e}}_2|$ of course follow the trend 
of $|\bm{p\cdot}\hat{\bm{\Omega}}|$ and peak at 0. This is because such spheroids show a strong alignment 
with the most contracting eigendirection $\hat{\bm{e}}_3$ as $\kappa \to 0$. While this is 
easier to understand for the $St = 0$ case, as reported in \cite{chevillard2013}, the problem 
of gravitationally settling inertial oblates is more subtle. In Figure~\ref{fig:pdfpdot}b 
we show the distributions for $|\bm{p\cdot}\hat{\bm{e}}_3|$ for a fixed $St\approx0.2$ and 
different aspect ratios. For nearly spherical oblates $\kappa = 0.95$, the distribution 
is \textit{close} to being uniform as one would expect. However, for truly anisotropic particles with 
$\kappa \leq 0.5$, we see clear evidence, even for such finite Stokes numbers, of the 
preferential alignment with the most contracting eigendirection with little or no dependence 
on the precise value of $\kappa$.

\begin{figure}
	\centering
	\includegraphics[width=1.0\columnwidth]{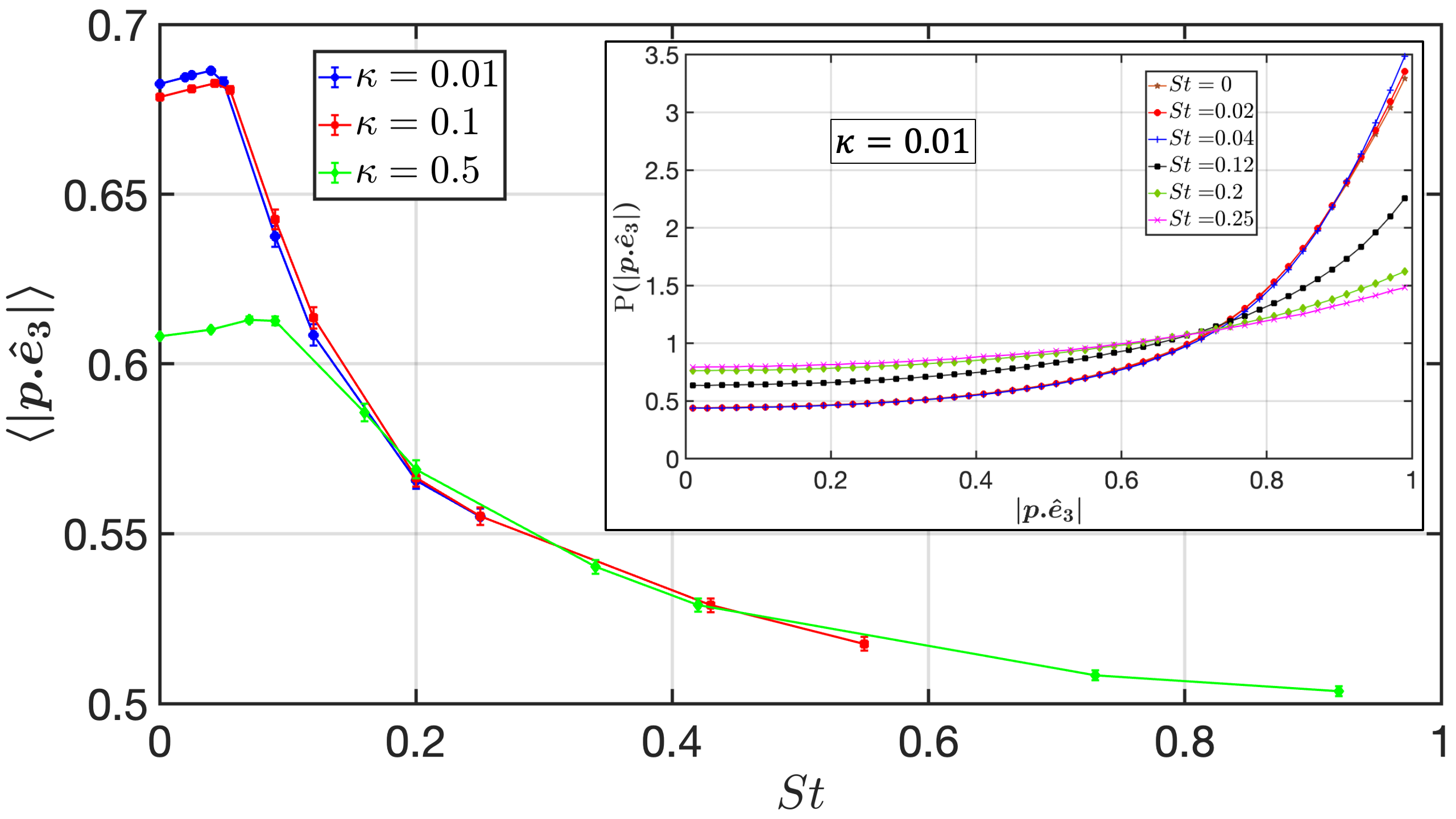}
	\caption{Mean values for the distributions of $|\bm{p\cdot}\hat{\bm{e}}_3|$ for all the aspect ratios as a function of Stokes number. For a given $\kappa$, the values decrease with increasing Stokes, and for small Stokes, the values depend on $\kappa$. The inset shows the probability distributions for $|\bm{p\cdot}\hat{\bm{e}}_3|$ for $\kappa=0.01$ for various Stokes numbers. The trend shown by the peak of the distribution is mirrored by the mean values.}
	\label{fig:meanvalues}
\end{figure}

Interestingly, our measurements for the distributions of  $|\bm{p\cdot}\hat{\bm{e}}_3|$ 
for different values of $St$ (for a given $\kappa$) show a major surprise (Figure~\ref{fig:meanvalues}, inset for 
$\kappa = 0.01$): 
The distributions themselves and in particular 
their peak values as $|\bm{p\cdot}\hat{\bm{e}}_3| \to 1$ show a sensitivity to the particle inertia, especially for 
the extreme shaped $\kappa \to 0$ particles, reminiscent of the measurements of $D_2$ and $\xi_1$ reported earlier.
While in the limit $St \to 0$ the distributions sharply peak at $|\bm{p\cdot}\hat{\bm{e}}_3| \to 1$, consistent 
with the tracer, non-sedimenting anisotropic particles studied in \cite{chevillard2013}, the extent 
of preferential alignment falls sharply as $\kappa \to 1$. 
We quantify this effect by calculating the mean $\langle|\bm{p\cdot}\hat{\bm{e}}_3|\rangle$ for such distributions 
and plotting their values as a function of $St$ for various $\kappa$ in Figure~\ref{fig:meanvalues}. 
For particles with similar inertia, this mean varies monotonically with $\kappa$ with the strongest 
alignment as $\kappa \to 0$. For our nearly spherical particles $\kappa = 0.95$, the mean value is essentially 0.5 
and, within error bars, independent of the $St$ numbers, as one would expect for $\kappa \to 1$. (For clarity, 
we do not show the corresponding plot for the near-sphere case $\kappa = 0.95$ where the distributions are essentially 
flat with $\langle|\bm{p\cdot}\hat{\bm{e}}_3|\rangle \approx 0.5$.)

In our model for sedimenting oblates, the  translational and
rotational dynamics are coupled in very non trivial ways. Our observations of the non-trivial 
shape and inertia dependence in measurements of pair statistics --- $D_2$ and $\xi_1$ --- as well 
as in single particle alignment statistics (Figure~\ref{fig:meanvalues}) leaves open the  
intriguing possibility of the two effects being related. Indeed, the mean $\langle|\bm{p\cdot}\hat{\bm{e}}_3|\rangle$ 
suggests that spheroids with large enough Stokes numbers sample the local flow geometry more \textit{uniformly}.
However, the precise relation between pair statistics and such single particle statistics is well beyond the 
scope of this manuscript but should serve as future direction of investigations.

%\begin{figure} 
%	\centering 
%	\begin{subfigure}[b]{0.49\columnwidth}
%		\includegraphics[width=\textwidth]{pdote3_kappa_p01_paper.png} 
%		\caption{}
%	\end{subfigure} 
%	\hfill 
%	\centering
%	\begin{subfigure}[b]{0.49\columnwidth}
%		\includegraphics[width=\textwidth]{meanvalues_pdote3_vs_Stokes_kappa.png} 
%		\caption{}
%	\end{subfigure} 
%	\caption{(a) Probability distributions for $|\bm{p\cdot}\hat{\bm{e}}_3|$ for $\kappa=0.01$ for various Stokes numbers. The peak of the distribution exhibits nonmonotonicity with respect to $St$.
%		(b)
%		Mean values for the distributions of $|\bm{p\cdot}\hat{\bm{e}}_3|$ for all the aspect ratios as a function of Stokes numbers. The values vary nonmonotonically with Stokes for a given $\kappa$, and increase with decreasing $\kappa$ for a given Stokes number.}
%	\label{fig:meanvalues} 
%	\end{figure}

\section{Conclusion}

Before we conclude, it is important to underline and discuss a possible shortcoming of our model which is beyond the scope of the present work. When considering approach rates of settling 
spheroids, it is tempting to ask if hydrodynamic interactions (HI) --- neglected in our work --- could play 
a dominant role in influencing the statistics that we report. On one hand, the question of 
HI is a delicate one and on the other the  neglect of such interactions is in 
the spirit of most of the work in this area. Given the relative infancy of studies of
sub-Kolmogorov spheroids in a turbulent flow, it is not surprising that, to the
best of our knowledge, no studies exist on the HI in such turbulent
suspensions unlike the case for sub-Kolmogorov spherical
particles~\citep{wang2005,wang2007JT,ayala2008effects,rosa2011,sticky,onishi2013}.
Briefly, while these studies do indicate that HI has an effect on the collision
efficiency, relative velocity and pair distribution, the difference between the
HI and the no HI cases turns out to be small. In fact, \cite{ayala2008effects}
find that the background turbulence severely limits the overall effect of HI.
Indeed, this conclusion stems from their observation that collision efficiency
remains more or less unchanged with and without HI. On a related note, a recent work~\cite{bragg2022} 
shows the limitations of a theoretical model based on HI for weakly
inertial particles in explaining the extreme clustering observed in the
experiments of~\cite{yavuz2018}. We stress again that these examples are
not to suggest that HI has no role in particle dynamics but perhaps they are
sub-dominant for sub-Kolmogorov particles in a turbulent flow. Indeed, for
turbulent suspensions, \cite{dhanasekaran2021} showed recently that the
importance of HI starts to emerge when spherical particles are within distances
of the order of the mean free path where the continuum theory breaks down. All of this
suggests that even for the simpler problem of turbulent suspensions of
spherical sub-Kolmogorov particles, neglect of HI is a reasonable approximation
when addressing questions of clustering dimensions and approach rates.

In summary, our work highlights the shape-sensitive nature of the two-particle statistics
for inertial spheroids when the competing effects of the turbulent and the
gravity-induced torques are factored in. In particular, through a systematic
and largely numerical study, we demonstrate the consequences
of this in the nature of clustering and approach rates for such particles.
This supplements several recent studies which have advanced our understanding
of the phenomenon of turbulent transport for anisotropic particles. In
particular, it complements the recent studies of collision
kernels~\citep{siewert2014collision,sheikh2022} in such systems by measuring
the approach rates of colliding particles.

The present work motivates several questions for future studies.
One of these relate to the alignment of the spheroid with the 
local geometry of the flow. \cite{chevillard2013} demonstrated that the Recent Fluid Deformation
Approximation (RFDA) --- a stochastic model for the velocity gradient
tensor proposed by \cite{chevillard2006} --- gives a good prediction for the
orientation statistics of the inertialess spheroids. It would be tempting, in the light of our results, 
to see how such models can be extended to the problem of finite inertia and their results  
tested against our DNS results which we report 
in this work. This could provide the theoretical underpinnings for what effects 
govern the dependence of preferential concentration and
the relative longitudinal velocity on the anisotropy of the particle. In fact,
one can also make theoretical predictions in future studies by developing a
model for the collision of spheroids settling in turbulence in a manner similar
to~\cite{Jiang_Xu_Zhao_2024}, who obtain the collision kernel for spheroids
settling in a quiescent ambient.  As we have already discussed, an obvious
restriction in our study is the smallness of $Re_s$, which is reflective of the
model used for the gravity-induced torque. Addressing the aforementioned
restriction is the next step to generalizing the results in our study. A first
attempt at the same was made by \cite{lopez2017} who proposed a theoretical
model, by using the framework developed in \cite{khayatcox1989}, to obtain the
settling velocity for slender fibers ($\kappa\gg1$) for $Re_s\sim O(1)$; they
reported a good match with their own experiments. Using their model, one can
suspend slender fibres in turbulence and study their orientational and
translational dynamics.  This will greatly help in expanding the parameter
space covered in the present study. Another potentially interesting study can
involve using asymmetric shapes settling in turbulence, in contrast to
spheroids (which are fore-aft symmetric). Recently, \cite{roy2019} calculated
the gravity-induced torque on freely settling asymmetric fibres for $Re_s\ll1$.
In the same spirit as in the present study, the aforementioned gravity-induced
torque can be implemented for the fibres settling in turbulent flow, which can
then be used to characterize their orientation dynamics and pair statistics.
Pursuing these questions will provide us with a more quantitative understanding
of the dynamics of anisotropic ice crystals in cold clouds as well as help in
addressing questions of mixed phase turbulent transport.

\section{Acknowledgements}

We thank G. Subramanian for introducing us to this subject and several discussions on this topic. 
We also thank H. Joshi for discussions. 
The simulations were performed on the ICTS clusters \emph{Tetris} and
\emph{Contra}. SSR acknowledges SERB-DST (India) projects STR/2021/000023 and
CRG/2021/002766 for financial support. We acknowledge the support of the DAE, Govt. of India, under project
no.  12-R\&D-TFR-5.10-1100 and project no. RTI4001.

\backsection[Declaration of Interests]{The authors report no conflict of interest.}
\backsection[Author ORCID]{
	P. Anand, https://orcid.org/0000-0001-6922-0894; S.S. Ray, https://orcid.org/0000-0001-9407-0007
}

%\appendix
%\section{}\label{appA}

%\begin{figure*}
%	\includegraphics[width=0.31\linewidth]{blank.png}
%	\includegraphics[width=0.31\linewidth]{blank.png}
%	\includegraphics[width=0.31\linewidth]{blank.png}
%	\caption{Representative plots of the pdfs of velocity differences $\delta v$, for $Fr = 0.05$, conditioned on the separation $r$ 
	%	for different values of $\kappa$ (see legend) for $St\sim 0.25$ and (a) $\eta\leq r\leq1.6\eta$, (b)$3\eta\leq r\leq 3.6\eta$, and (b) $5\eta\leq r\leq5.6\eta$. The insets show corresponding 
	%	plots for the zero gravity ($Fr = \infty$) case.}
%	\label{fig:vpdfVsStokes}
%\end{figure*}

\bibliographystyle{jfm}
\bibliography{references}

\end{document}